\documentclass[a4paper,aps,pra,twocolumn,10pt,superscriptaddress,longbibliography]{revtex4-1}
\usepackage{stmaryrd}
\usepackage{}
\usepackage{amssymb}
\usepackage{mathrsfs}
\usepackage{color}
\usepackage{graphicx}
\usepackage{dcolumn}
\usepackage{bm}
\usepackage{hyperref}
\usepackage{bbm}
\usepackage{amsmath}
\usepackage{verbatim}

\begin{document}

\preprint{APS/123-QED}

\title{Entangling and squeezing atoms by weak measurement}
\author{Mingfeng Wang}
\affiliation{Department of Physics, State Key Laboratory of Surface Physics and Key Laboratory of Micro and Nano Photonic Structures (Ministry of Education), Fudan University, Shanghai 200433, China}
\affiliation{Department of Physics, Wenzhou University, Zhejiang 325035, China}
\author{Weizhi Qu}
\affiliation{Department of Physics, State Key Laboratory of Surface Physics and Key Laboratory of Micro and Nano Photonic Structures (Ministry of Education), Fudan University, Shanghai 200433, China}%
\author{Han Bao}
\affiliation{Department of Physics, State Key Laboratory of Surface Physics and Key Laboratory of Micro and Nano Photonic Structures (Ministry of Education), Fudan University, Shanghai 200433, China}%
\author{Pengxiong Li}
\affiliation{Department of Physics, State Key Laboratory of Surface Physics and Key Laboratory of Micro and Nano Photonic Structures (Ministry of Education), Fudan University, Shanghai 200433, China}%
\author{Yanhong Xiao}
\email{yxiao@fudan.edu.cn}
\affiliation{Department of Physics, State Key Laboratory of Surface Physics and Key Laboratory of Micro and Nano Photonic Structures (Ministry of Education), Fudan University, Shanghai 200433, China}%
\affiliation{Collaborative Innovation Center of Advanced Microstructures, Nanjing 210093, China}
\date{\today}
\date{\today}
\date{\today}

\begin{abstract}
A weak measurement approach is proposed to entangle and squeeze atoms. We show that even for very small coupling strength between light and atoms, one can achieve large squeezing unattainable with normal measurement-based squeezing. Post-selecting the photons interacting with spins via a weak off-resonant quantum nondemolition interaction (QND) in a state nearly orthogonal to its initial state can strengthen the entanglement among elementary spins and thus enhance spin squeezing. The squeezing process is probabilistic but the created state is unconditionally determined. Further control of the post-selection parameter and the detection process can transform the QND-like spin squeezing into the one-axis-twisting or even two-axis-twisting type.
\\
\begin{description}
\item[PACS numbers]
03.67.Bg, 42.50.Dv, 03.65.Ud
\end{description}
\end{abstract}

\maketitle


 \makeatletter
    \newcommand{\rmnum}[1]{\romannumeral #1}
    \newcommand{\Rmnum}[1]{\expandafter\@slowromancap\romannumeral #1@}
    \makeatother

  Creating multipartite entangled spin states is of general interest to quantum metrology  \cite{PhysRevA.46.R6797,PhysRevLett.82.4619,Nautre1} and quantum information science \cite{PhysRevLett.85.5643,RevModPhys.82.1041, NATRUE3}. In particular, spin squeezed states (SSS) --- a special category of entangled spin states~\cite{PhysRevA.47.5138} whose total-spin fluctuations along a direction are smaller than the sum of the individual-spin fluctuations, can enable atomic clocks~\cite{1367-2630-12-6-065032,PhysRevLett.104.250801} and magnetometers~\cite{PhysRevLett.109.253605,PhysRevLett.104.093602} with sensitivities surpassing the standard quantum limit. To date, several mechanisms have been investigated to produce SSS, including nonlinear interactions among the individual atoms either directly or mediated by light \cite{PhysRevLett.85.3991,PhysRevA.68.043622,PhysRevLett.83.2274,PhysRevA.66.022314,PhysRevLett.94.023003,PhysRevLett.105.193602,PhysRevA.62.063812,
PhysRevLett.110.120402,PhysRevLett.104.073602}, quantum-state transfer from squeezed light to atomic ensemble \cite{PhysRevLett.84.4232,PhysRevLett.83.1319,PhysRevLett.89.057903,PhysRevLett.88.070404,PhysRevLett.87.170402,PhysRevLett.90.030402,PhysRevLett.107.013601,
PhysRevLett.106.010404}, and QND projective measurements of collective spin \cite{NATRUE33,PhysRevA.60.4974,PhysRevA.60.2346,PhysRevLett.85.1594,PhysRevA.65.061801,PhysRevLett.102.033601,PhysRevLett.110.163602}. Among them, QND measurement \cite{EPL1} is most general as it is experimentally realizable in many atomic systems, ranging from cold atoms \cite{PhysRevLett.102.033601,PhysRevLett.106.133601} to room temperature vapors \cite{naturephysics66}.
It usually involves off-resonant atom-light interaction and produces entanglement between light and the atomic spin. A projection measurement of light then maps the atoms into a spin-spin entangled state, yielding a conditional quantum-noise reduction $\propto 1/(1+\kappa^2)$, where $\kappa$ is the coupling strength between light and atoms which can be increased by either applying a cavity or using an elongated cooled atomic ensemble. However, many precision measurement systems in free space, in particular those with large atom numbers making the most sensitive atomic sensors \cite{nature66}, have inherently low coupling strength $\kappa$, for instance, $\kappa\simeq 1$ in hot atomic vapors \cite{SSA2018} and $\kappa\simeq 0.6$ in cold atomic vapors \cite{PhysRevLett.110.163602}, where large squeezing is unattainable with currently available QND techniques.

In this paper, we propose a novel scheme to entangle and squeeze atoms using weak measurement (WM) \cite{PhysRevLett.60.1351}. The scheme gives enhanced squeezing compared to the normal QND method especially in the weak coupling regime. The WM, a quantum measurement protocol first introduced by Aharonov \emph{et al.} \cite{PhysRevLett.60.1351}, investigates a situation where the coupling between the system and the probe is weak. Appropriately pre- and post-selected system states, i.e., nearly orthogonal, yield a counterintuitive result --- the distribution of measured values can be dramatically amplified and thus lie significantly outside the range of eigenvalues of the observable operator. The ability of WM to amplify tiny physical effects has led to numerous applications, including the measurements of small frequency shifts \cite{PhysRevA.82.063822}, the amplification of small phase \cite{PhysRevA.82.011802} and optical nonlinearities \cite{PhysRevLett.107.133603}. We here show that the WM can be used to produce and strengthen the entanglement among elementary spins. Unlike traditional conditional QND squeezing, the proposed WM-based protocol is \emph{unconditional}, i.e., the resulted quantum spin distribution always has zero mean and is squeezed along the same direction, even independent of the coupling strength. Furthermore, by properly choosing the post-selection parameter and using a series of optical probe pulses, the QND-type squeezing can be converted into one-axis twisting (OAT) and two-axis twisting (TAT) squeezing, leading to a quantum-noise reduction that scales \emph{exponentially} with the coupling strength.

\emph{Squeezing generation by weak measurement}. We consider an atomic ensemble consisting of $N_A$ four-level atoms in the ground states $\left|\uparrow\right\rangle$, $\left|\downarrow\right\rangle$ (with $x$ as the quantization
axis) interacting off-resonantly with a light beam propagating along the $z$-axis, as shown in Fig. 1. If the light is tuned far from resonance, one may adiabatically eliminate the exited states and obtain the well-known Faraday rotation (FR) Hamiltonian \cite{PhysRevLett.85.5643,RevModPhys.82.1041, NATRUE3}: $H = \hbar\chi {S_z}{J_z}$,
where $\chi$ is the coupling constant, $S=(S_x,S_y,S_z)$ denotes the Stokes vector, and $J=(J_x,J_y,J_z)$ stands for the collective spin operators for the ground states of atoms. They obey the angular momentum  commutation relations $[{V _j},{V _k}] = i{\varepsilon _{jkl}}{V _l}$ with $V  \in \{ S,J\}$ and $j,k,l \in \{ x,y,z\}$. In particular, ${S_z} = \frac{1}{2i}( {a_x^\dag {a_y} - a_y^\dag {a_x}} )$ and ${J_z} = \frac{1}{2i}\sum\nolimits_{j = 1}^{N_A} {(\left| {{ \downarrow _j}} \right\rangle \langle { \uparrow _j}| - \left| {{ \uparrow _j}} \right\rangle \langle { \downarrow _j}|)}$, where $a_x(a_y)$ is the annihilation operator of $x(y)$-polarized light. We assume the light field contains a strong $x$-polarized component and a weak $y$-polarized component, such that one can treat the $x$-mode operators $a_x^\dag,a_x$ as classical $c$-numbers and the $y$-mode field as the quantum field which is relevant here. Under this condition, the
time evolution operator can be written as  $U =\text{exp} (-i{\kappa _0}{P_L}{J_z})$, where we have defined a new constant $\kappa_0=\chi\sqrt{N_{ph}/2}$ with $N_{ph}$ the photon number of the light mode and the canonical operators  $X_L=(a_y+a_y^\dag)/\sqrt{2},P_L=-i(a_y-a_y^\dag)/\sqrt{2}$ for $y$-polarized light, satisfying $[X_L,P_L]=i$.
\begin{figure}[t]
\centering
\includegraphics[scale=0.3]{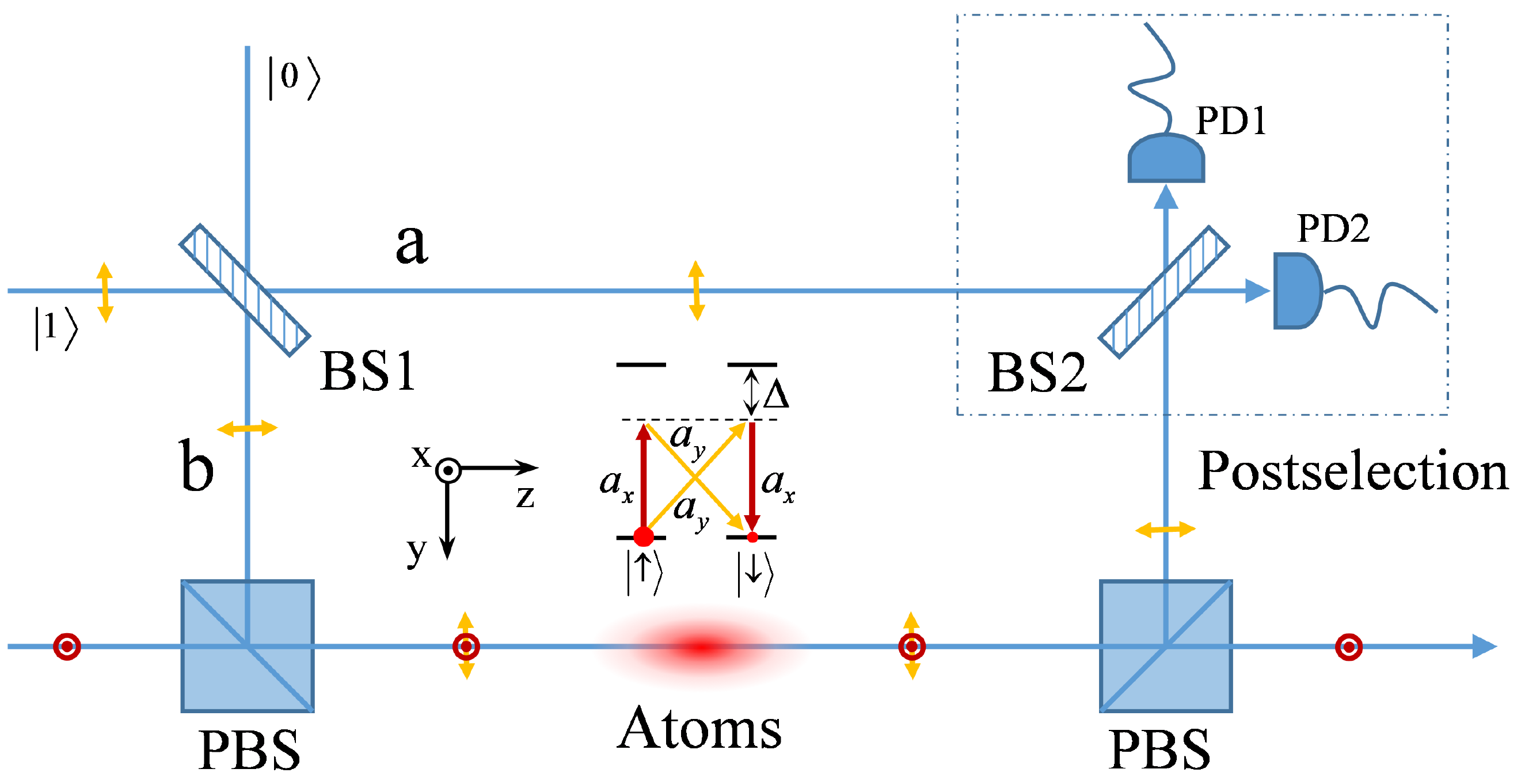}
\caption{ Schematics setup of the proposed scheme. A $y$-polarized single photon is sent through a beam splitter BS1. One of the output mode b is then combined with a strong $x$-polarized coherent light on a PBS, which is then injected into the sample to experience the off-resonant Faraday rotation interaction (QND type). After the interaction, the probe light is
guided into another PBS to separate the two polarization modes, where the $y$-polarized photons recombine at BS2 with the output mode a of BS1, and detected at upper detector PD1 and lower detector PD2.}
\end{figure}

We begin by revisiting the traditional QND-based spin squeezing. The atoms are initially prepared in the coherent spin state (CSS) $\left| {{ \Psi_A }} \right\rangle_{{in}}  = {\left|  \uparrow  \right\rangle ^{ \otimes {N_A}}}$ along $x$, and the relevant $y$-polarization component of the input $x$-polarized light can be written as a vacuum state $|\varphi_L\rangle=|0_L\rangle$. After the QND interaction, a homodyne detection of the $X_L$ component is performed, leading to a random measurement result $X_L=x_L$, for which $\langle x_L|P_L|0_L\rangle=\langle x_L|0_L\rangle ix_L$, and $\langle x_L|P_L^2|0_L\rangle=\langle x_L|0_L\rangle(1-x_L^2)$. For $\chi\ll 1$, the atom ensemble is projected into a state $|\Psi_A\rangle_{out}=\langle x_L|U|\varphi_L\rangle\left| {{ \Psi_A }} \right\rangle_{{in}} \approx \langle x_L|0_L\rangle [\left| {\uparrow}\right\rangle^{\otimes {N_A}}-i\frac{\kappa}{\sqrt{2N_A}} x_L \left|\downarrow\right\rangle_{SSF}+ \frac{\kappa ^2}{4{N_A}}(1-x_L^2)\left| {\downdownarrows} \right\rangle_{PSF}]$, where we have kept only up to the second order in $U$, and defined the single-spin-flipped state $\left| {{ \downarrow }} \right\rangle_{{SSF}}=2i{J_z}\left| {{ \Psi_{A} }} \right\rangle_{{in}} = \sum\nolimits_{j = 1}^{{N_A}} {\left| {{ \downarrow _j}} \right\rangle \left|  \uparrow  \right\rangle _{ \ne j}^{ \otimes \left( {{N_A} - 1} \right)}}$ as well as the pairwise-spin-flipped state $\left| {{ \downdownarrows }} \right\rangle_{{PSF}}=(N_A-4J_z^2)\left| {{ \Psi_{A} }} \right\rangle_{{in}} = \sum\nolimits_{j,k = 1}^{{N_A}} {\left| {{ \downarrow _j}{ \downarrow _k}} \right\rangle \left|  \uparrow  \right\rangle _{ \ne j,k}^{ \otimes \left( {{N_A} - 2} \right)}}$, and also introduced $\kappa^2=\kappa_0^2N_A/2=\eta\alpha_0$, with the on-resonance optical depth $\alpha_0=N_A\sigma/A$ and the scattered photon number per atom $\eta=N_{ph}\sigma\Gamma^2/A\Delta^2$  \cite{PhysRevLett.85.5643,RevModPhys.82.1041, NATRUE3}, where $\sigma$ is the resonant light scattering cross section of a single atom, $A$ is the optical beam cross section, $\Gamma$ is the spontaneous decay rate, and $\Delta$ the detuning from resonance. We note that the second term $\left| {{ \downarrow }} \right\rangle_{{SSF}}$ contributes an average mean value (proportional to $x_L$) of $J_z$, which is often different for each detection and makes the protocol conditional; the third term $\left| {{ \downdownarrows }} \right\rangle_{{PSF}}$ is the essence of spin squeezing \cite{PhysRevLett.109.173603}, but has small weight for weak coupling.

Here, we propose to use the WM approach to enhance the spin-spin entanglement and thus the collective spin squeezing. As shown in Fig. 1, a $y$-polarized single photon is sent through a beam splitter (BS1), with reflectivity $r$ and transmissivity $t=\sqrt{1-r^2}$. As a result, the spatial modes $a$ and $b$ of the two output ports of BS1 are prepared in the entangled state $\left| {{\varphi _L}} \right\rangle  = r\left| {{0_a}} \right\rangle \left| {{1_b}} \right\rangle  + t\left| {{1_a}} \right\rangle \left| {{0_b}} \right\rangle $. Next, the spatial mode $b$ and a strong $x$-polarized beam are combined by a PBS to produce a single output beam with independent $x$ and $y$ polarizations. The output beam is then injected into the atomic ensemble to experience the FR interaction, and after the interaction the probe light is guided into another PBS to separate the two polarization modes. The $y$-polarized photons in $a$ and $b$ are then recombined at BS2, with reflectivity $r'$ and transmissivity $t'=\sqrt{1-r'^2}$. The recombined photons are finally detected at upper detector PD1 and lower detector PD2. If PD1 detects `no photon' and PD2 detects `one photon', which corresponds to post-selecting the optical fields in a state, $\left| {{\varphi' _L}} \right\rangle  = r'\left| {{0_a}} \right\rangle \left| {{1_b}} \right\rangle  - t'\left| {{1_a}} \right\rangle \left| {{0_b}} \right\rangle $, collapsing the collective spin into the state
\begin{eqnarray}
|{\Psi _A}{\rangle _{out}} &=& \left\langle {{{\varphi '}_L}} \right|U\left| {{\varphi _L}} \right\rangle {\left| {{\Psi_A}} \right\rangle _{in}} \nonumber\\&\approx& \langle {{{\varphi '}_L}}| {{\varphi _L}} \rangle
 \left[ {{{\left|  \uparrow  \right\rangle }^{{\otimes N_A}}} + \frac{{{A_w}{\kappa ^2}}}{{4{N_A}}} \left| {{ \downdownarrows }} \right\rangle_{{PSF}}} \right]\label{eeq4}.
\end{eqnarray}
where the weak value (post-selection parameter) is
\begin{eqnarray}
{A_w} = \frac{{\left\langle {{{\varphi '}_L}} \right|P_L^2\left| {{\varphi _{L}}} \right\rangle }}{{\langle {{\varphi '}_L}|{\varphi _L}\rangle }} = \frac{{rr'}}{{rr' - tt'}} + \frac{1}{2}\label{eq45}. \end{eqnarray}
We note that (i) if the postselected state $|{\varphi '}_L\rangle$ is near orthogonal to the input state $|{\varphi }_L\rangle$, $\langle{\varphi '}_L|\varphi _L\rangle$ is very small, leading to a large $A_w$, which therefore enhances the coupling strength $\kappa$ effectively by a large factor $\sqrt{A_w}$, as seen from Eq.(1). (ii) the cancellation of the single-spin-flipped term $\left| {{ \downarrow }} \right\rangle_{{SSF}}$ ensures that the created squeezed state has a deterministic mean value of zero, indicating that the squeezing process is unconditional.
The probability of getting such a state is given by $P=|\langle{\varphi '}_L|\Psi _{A}\rangle_{out}|^2 \propto |\langle{\varphi '}_L|\varphi _L\rangle|^2 =(rr' - tt')^2$, showing that the price for the enhancement in squeezing is a reduction in the success probability.
\begin{figure*}
[t]\resizebox{17.5cm}{!}
{\includegraphics[scale=1.3]{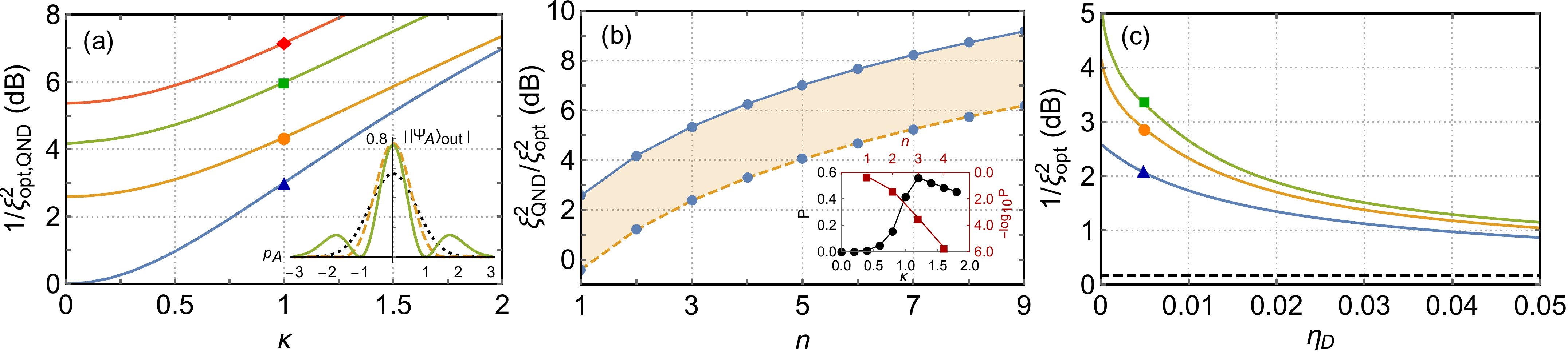}}
\caption{\label{fig:figure2} (Color online) (a) Performance of the QND (blue line with triangle) and the proposed WM spin squeezing protocols versus the coupling strength $\kappa$, for detection times $n=1$ (orange line with circle), $n=2$ (green line with square), and $n=3$ (red line with diamond). Inset: Probability distributions of the spin state in the $P_A$ basis for $|0_A\rangle$ (dotted line), $|o_2\rangle$ (dashed line, $A_w=74.3$), and $|e_2\rangle$ (solid line, $A_w=200$), where we take $\kappa=0.1$. (b) The amount of spin-squeezing enhancement of the WM protocols compared to QND, versus $n$ for coupling strength $\kappa\rightarrow 0$ (solid line) and $\kappa\rightarrow\infty$ (dashed line). Inset:  The probability of success $P$ vs $\kappa$ for $n=1$ (black polygonal line), and the probability of success $P$ vs $n$ (red polygonal line) for $\kappa=1.5$. (c) Maximal squeezing versus $\eta_D=1-\mathcal {E}_D$ with $\mathcal {E}_D$ the photon detection efficiency, for $\kappa=0.2$ and detection times $n=1$ (blue line with triangle), $n=2$ (orange line with circle), and $n=3$ (green line with square). The dashed line represents the squeezing produced
by the normal QND protocol.}
\end{figure*}

 Now we quantitatively show the amount of squeezing created by this WM approach. We employ the Holstein-Primakoff approximation \cite{PhysRev.58.1098}, and assume that the atomic ensemble is strongly polarized with large $N_A$ so that the small deviations from perfect polarization direction during interaction can be neglected. One can then define the canonical variables for atoms $X_A=J_y/\sqrt{N_A/2}$ and $P_A=J_z/\sqrt{N_A/2}$, satisfying $[X_A,P_A]=i$ and $X_A^2/2+P_A^2/2=n_A+1/2$, where $n_A$ denotes the atomic number operator \cite{PhysRevLett.91.060401}. With this notation we can rewrite the unitary evolution as $U=\exp(-i\kappa P_LP_A)$ and the CSS in terms of $p_A$ eigenstates as $|{\Psi _A}{\rangle_{in} } = \int dp_A\text{exp}[-p_A^2/2]|p_A\rangle$ (hereafter we omit normalization constants for simplicity), and, after taking into account effects of the higher-order terms of $U$, we have
\begin{eqnarray}
|{\Psi _A}{\rangle _{out}} &=& \left( {1 + \sum\limits_{n = 1}^\infty  {\frac{{{{(i\kappa )}^{2n}}\langle {{\varphi '}_L}|P_L^{2n}\left| {{\varphi _L}} \right\rangle }}{{(2n)!\left\langle {{\varphi '}_L}\right| {{\varphi _L}} \rangle }}} P_A^{2n}} \right)|{\Psi _A}{\rangle _{in}}\nonumber\\
 &=& \int {d{p_A}\left( {1 - {{\tilde A}_w}{\kappa ^2}p_A^2} \right){e^{ - \frac{{p_A^2}}{2\xi_s^2}}}\left| {{p_A}} \right\rangle }\label{eq33}
\end{eqnarray}
where $\tilde A_w=A_w/2-1/4$ and $\xi_s^2=1/(1+\kappa^2/2)$.  For this state, one can calculate the atomic variance $(\Delta {P_A})^2$ and thus the squeezing parameter $\xi^2  = {(\Delta {P_A})^2}/{(\Delta {P_{{A}}^{CSS}})^2}$. Optimizing $\xi^2$ with respect to $A_w$, we get $\xi^2_{opt} =(3-\sqrt{6}) \xi_s^2=0.55 \xi_s^2$ for $A_w=4(3-\sqrt{6})/(3\kappa^2)+(15-4\sqrt{6})/6$. compared to the traditional QND whose squeezing parameter is $1/(1+\kappa^2)$, this schemes can enhance the amount of squeezing especially in the weak coupling regime. For instance, as $\kappa\rightarrow 0$, QND gives nearly no squeezing, and the squeezing enhancement by our WM method reaches the limit of about $2.6$ dB (given by the coefficient 0.55). In Fig. 2(a), we plot the squeezing created by this WM protocol and the QND protocol, which verifies that the enhancement is more significant in the weaker coupling regime and decreases for larger $\kappa$.

An interesting aspect of this scheme is that a non-Gaussian entangled state can be created (at large weak values $A_w$) which also has applications in quantum metrology~\cite{PhysRevLett.110.163604}, even though such entangled state does not favor squeezing and is precisely the reason for the above $2.6$ dB limit in squeezing enhancement. The physics can be better understood if we express the state of Eq. (\ref{eq33}) in terms of the atomic number state $|n_A\rangle$: $|{\Psi _A}{\rangle_{out} } =(1-\kappa^2\tilde A_w\xi^2_s/2)|0_A\rangle_\zeta+\kappa^2\tilde A_w\xi^2_s/2|2_A\rangle_\zeta$, where $|n_A\rangle_\zeta=S(\zeta)|n_A\rangle$ denotes the squeezed atomic number state, with the squeezing operator $S(\zeta)=\exp[-\zeta(a_A^2-{a_A^\dag}^2)/2]$, where $\zeta=\ln\xi_s^2$ and $a_A$ is the atomic annihilation operator. (i) For small $\kappa$, $\zeta$ is nearly zero, and the squeezer $S(\zeta)$ has negligible contribution to squeezing, i.e., $|n_A\rangle_\zeta\simeq |n_A\rangle$. In this regime, by choosing suitable $A_w$ in post-selection, one can collapse the atomic state from the initial vacuum state $|0_A\rangle$ to a superposition state $|o_2\rangle=(1/2+1/{\sqrt{3}})^{1/2}|0_A\rangle+(1/2-1/{\sqrt{3}})^{1/2}|2_A\rangle$, and even to an equal superposition state  (for larger $A_w$) $|e_2\rangle=(|2_A\rangle-|0_A\rangle)/\sqrt{2}$. The state $|o_2\rangle$ has the narrowest probability distributions in the $P_A$ basis [shown as the dashed line in the inset of Fig. 2(a)], corresponding to the scenario of best enhancement in squeezing compared to QND. For much larger $A_w$, $|o_2\rangle$ approaches $|e_2\rangle$, which deviates from the normal Gaussian state [solid line in the inset of Fig. 2(a)] and causes a diffusion of the probability distributions and hence increased variances of $P_A$. (ii) For larger $\kappa$, the squeezer $S(\zeta)$ contributes to squeezing by $\xi_s^2$.  Note that the success probability $P$ increases rapidly with the coupling strength $\kappa$ [see the inset of Fig. 2(b)] and reaches its maximum value $P=0.58$ at $\kappa=1.1$. Such characteristics are attractive since the relatively large success probability results in near-deterministic unconditional squeezing, and more importantly, many free-space atomic systems, such as warm vapor cell \cite{0953-4075-41-22-223001} and cold atoms \cite{PhysRevLett.102.033601}, work well for precision measurement with this small coupling-strength.

\emph{Multi-detection scheme and one(two)-axis-twisting squeezing.} As discussed above, the enhancement in squeezing over QND is moderate and determined by the state $|o_2\rangle$. However, one can obtain much higher enhancement by employing the multi-detection scheme and constructing higher-dimensional superposition state $|o_{2n}\rangle=\sum^n_{j=0}c_j|(2j)_A\rangle$ with $n>1$ and $c_j$ the normalization constant, which is realized by breaking the strong $x$-polarized light pulse into $n$ subpulses. Each subpulse co-propagates with a $y$-polarized superposition state described above, and then together experience the FR interaction and subsequent single photon detections. If all the detections succeed, the spin state is collapsed into:
\begin{eqnarray}
|{\Psi _A}{\rangle _{out}} =\int {d{p_A}\prod\limits_{j = 1}^n {\left( {1 - {{ {{\tilde A_w} } {\vartheta _j^2}{\kappa ^2}p_A^2}}} \right)} }
{e^{ - \frac{{{}p_A^2}}{2\xi_s^2}}}\left| {{p_A}} \right\rangle,\label{eq4}
\end{eqnarray}
where $\vartheta _j<1$ relates to weight of each subpulse in photon number and satisfies $\sum\nolimits_{j = 1}^n {\vartheta _j^2}  = 1$. Corresponding to this state, in Fig. 2(a) we also plot the optimal squeezing of the
protocol (optimized with respect to $\vartheta _j$) for varying $\kappa$ under different detection times $n=2,3$, which indicates that the improvement in the enhancement can be significant. Figure 2(b) shows the enhancement as a function of detection times $n$, for $\kappa\rightarrow 0$ and $\infty$, indicating that, when $n>1$, the QND squeezing are completely surpassed by the WM squeezing, and an enhancement of $9$ dB is obtainable at $n=9$.

Although of solely theoretical value, interesting results appear when we set $\vartheta_j^2=1/n$ (which corresponds to the case of $n$ equal subpulse) with $n\rightarrow\infty$ and further manipulate the weak value $A_w$. We have
\begin{eqnarray}
|{\Psi _A}{\rangle _{out}} \simeq\int {d{p_A}{e^{ - \frac{{{A_w}{\kappa ^2}p_A^2}}{2}}}{e^{ - \frac{{p_A^2}}{2}}}\left| {{p_A}} \right\rangle }\label{eq5}.
\end{eqnarray}
For this state, if $A_w=1$, the output state is exactly the same as the state created by QND scheme \cite{PhysRevLett.91.060401}. Therefore, any $A_w>1$ will improve the performance of squeezing. Note that $A_w$ can be a positive, negative and even complex number \cite{PhysRevA.76.044103}. Consider the particular case when $A_w$ is imaginary \cite{imageaw}, i.e., $A_w \rightarrow iA_{w}$, then, according to Eq. (\ref{eq5}), the net effect of the multi-detection is equivalent to a unitary transformation $U_{OAT}=e^{-iA_{w}\kappa^2p_A^2/2}$, which is exactly the well-known OAT transformation \cite{PhysRevA.47.5138}, yielding a squeezing $\xi^2_{OAT}\simeq1/A_{w}^2\kappa^4$ that is a quadratic improvement of the QND protocol. Further improvement is possible if one can transform OAT into TAT \cite{PhysRevLett.107.013601}. To do so, the subpulses described above are sent through the sample alternately along the $x$ and $y$ directions. The subsequent (successful) post-selection measurements result in the effective transformation $U_{TAT}\simeq(1+iA_w\kappa^2X_A^2/n)^{n/2}(1-iA_w\kappa^2P_A^2/2)^{n/2}\simeq\exp[-iA_w\kappa^2(P_A^2-X_A^2)]$, where the post-selection processes are chosen such that the $y$-direction yields $A_w\rightarrow -iA_w$ while the $z$-direction generates $A_w\rightarrow iA_w$. The transformation $U_{TAT}$ creates a squeezing $\xi_{TAT}^2=e^{-A_w\kappa^2}$, indicating that the spin fluctuations are now shrunken exponentially.

\emph{General considerations.} A major challenge in implementing the multi-detection protocol is that its success odds decreases dramatically with $n$, as shown in the inset of Fig 2(b). To increase the odds, we suggest utilizing the non-maximally entangled NooN state $|\varphi_{L}\rangle=r|0_a\rangle|m_b\rangle+t|m_a\rangle|0_b\rangle$ \cite{PhysRevLett.85.2733} as the input state and, correspondingly, the state $|\varphi_{L}'\rangle=r'|0_a\rangle|m_b\rangle-t'|m_a\rangle|0_b\rangle$ as the post-selection state, which, for weak coupling, leads to the same state as Eq. (\ref{eeq4}) and yields the $m$-photon weak value $A_{w,m}=2m\tilde A_w$ for large $m$. Since the success odds of the multi-detection protocol is $P_n\propto 1/(2\tilde A_w)^{2n}$, the use of multi-photon NooN state can thus greatly increase the odds, that is $P_n\rightarrow m^{2n}P_n$.

We next show that the proposed scheme can also be extended to the case of coherent-state input $|\alpha\rangle$, which is easier to implement than the single-photon Fock state \cite{nature60}. Then, the initial state after BS1 is $|\varphi_L\rangle=|r\alpha\rangle_a|t\alpha\rangle_b$. If the light state is successfully post-selected in the state $|\varphi_L'\rangle=\sqrt{2}r't'(|2_a\rangle|0_b\rangle-|0_a\rangle|2_b\rangle)+(t'^2-r'^2)|1_a\rangle|1_b\rangle$, the spin state collapses into $|\Psi_A\rangle_{out}\simeq\langle\varphi_L'|\varphi_L\rangle(1-i\kappa A_0 P_A-\frac{1}{2}\kappa^2 A_{w,\alpha} P_A^2)|\Psi_A\rangle_{in}$ with $A_{w,\alpha}=(1-i\alpha A_0-2r_0'/\alpha^2)/2$, where we have assumed $r=t=1/\sqrt{2}$, $A_0=i(r_0'-\alpha^2/2+1)/\alpha$ and $r_0'=2r't'/(t'^2-r'^2)$. If $\alpha=\sqrt{2(r_0'+1)}$, we get $A_0=0$ and $A_{w,\alpha}=1/[2(r_0'+1)]$. The spin state obtained is exactly the same as the state in Eq. (\ref{eeq4}), but with $A_w$ replaced by $A_{w,\alpha}$. Large values of $A_{w,\alpha}$ are obtainable when $r_0'$ approaches $-1$.

So far, we have ignored the decoherence processes where the transverse $J_y$ and $J_z$ components decay with coefficient $\eta$ and the $y$-polarized quantum field is absorbed with a ratio $\epsilon=\eta N_A/N_{ph}\simeq\eta$ for $N_A\sim N_{ph}$ \cite{PhysRevA.70.044304}. For the case of weak coupling, such losses can be neglected since $\eta$ is small. Another inevitable imperfection is the inefficiency of the photodetectors. The single-photon detectors usually have a detection efficiency $\mathcal {E}_D<1$, which can be modeled as a sequence of (virtual) beam splitters with reflectivity $\eta_D=1-\mathcal {E}_D$ followed by an idealized detector \cite{PhysRevA.72.033822}. Figure 2(c) shows how our protocols perform for different $\eta_D$, from which we can see that, for a feasible value $\eta_D=5\%$ \cite{sp1,sp2}, the squeezing produced is still well above the QND bound on the spin squeezing, while higher-efficiency detectors
are required to make the multi-detection scheme work well.

\emph{Conclusion.} We proposed a novel scheme to prepare atoms in an entangled or squeezed state, by using weak measurement. The scheme is probabilistic but unconditional, and can have significant enhancement in squeezing over traditional QND scheme, especially in the weak coupling regime. The QND-type interaction can be even turned into one-axis-twisting and two-axis-twisting type, simply by controlling the post-selection parameters. This scheme can be used as a general state-preparation method and should be useful in the context of entanglement study and quantum metrology.

We thank Vladan Vuleti\'{c} for enlightening discussions and acknowledge support from National Key Research Program of China under Grant No. 2016YFA0302000 and NNSFC under Grants No. 11504273 (M.Wang), 91636107, 61675047.

\bibliographystyle{apsrev4-1}
\bibliography{ref}

\begin{thebibliography}{57}
\bibitem [{ima()}]{imageaw}%
  \BibitemOpen
  \href@noop {} {\bibinfo  {journal} {If the output optical state is
  postselected to, e.g., the superposition state $\left| {{\varphi' _L}}
  \right\rangle = r'\left| {{0_a}} \right\rangle \left| {{1_b}} \right\rangle -
  e^{-i\pi/6}t'\left| {{1_a}} \right\rangle \left| {{0_b}} \right\rangle $
  (which can be realized by placing a wave plate to one arm of the BS2), one is
  able to obtain an imaginary post-selection parameter
  $A_w=i/(2\sqrt{3}-4rr'/tt')$}\ }\BibitemShut {NoStop}%
\end{thebibliography}
\end{document}